\documentclass[12pt]{article}
\input{epsf}
\usepackage{graphicx}
\def\AJ{{\it Astroph. J.} }

\def\GRG{{\it Gen. Relativity and Gravitation} }

\def\PL{{\it Phys. Lett.} }
\def\PR{{\it Phys. Rev.} }

\def\frac#1#2{{\textstyle{{#1}\over {#2}}}}

\def\lsim{\mathrel{\rlap{\lower4pt\hbox{\hskip1pt$\sim$}}
    \raise1pt\hbox{$<$}}}
\def\gsim{\mathrel{\rlap{\lower4pt\hbox{\hskip1pt$\sim$}}
    \raise1pt\hbox{$>$}}}
\def\sqr#1#2{{\vcenter{\vbox{\hrule height.#2pt
         \hbox{\vrule width.#2pt height#1pt \kern#1pt
         \vrule width.#2pt}
         \hrule height.#2pt}}}}

\def\beq{\begin{equation}}
\def\eeq{\end{equation}}
\def\beqa{\begin{eqnarray}} 
\def\eeqa{\end{eqnarray}}

\def\laq{\raise 0.4 ex \hbox{$<$}\kern -0.8 em\lower 0.62 ex\hbox{$\sim$}}
\def\gaq{\raise 0.4 ex \hbox{$>$}\kern -0.7 em\lower 0.62 ex\hbox{$\sim$}}
\begin{document}
\titlepage

\begin{center}
{\bf Generalized Chaplygin Gas Model: \\
Dark Energy - Dark Matter Unification and CMBR
Constraints}\footnote{Essay selected for an honorable mention by the
  Gravity Research Foundation, 2003.}
\vglue 1.5cm
{M. C. Bento\footnote{Also at CFIF, Instituto Superior T\'ecnico, Lisboa.
  Email address: bento@sirius.ist.utl.pt}, O. Bertolami\footnote{Also at CFNUL,
Universidade de Lisboa. Email address: orfeu@cosmos.ist.utl.pt},
 A. A. Sen\footnote{Also at CENTRA, Instituto Superior T\'ecnico, Lisboa.
  Email address: anjan@x9.ist.utl.pt}
\\}
\bigskip
{\it Instituto Superior T\'ecnico,
Departamento de F\'\i sica,\\}
\medskip
{\it Av.\ Rovisco Pais 1, 1049-001 Lisboa, Portugal\\}

\vglue 1cm

\end{center}
\baselineskip=20pt

\centerline{\bf  Abstract}
\vglue 1cm
\noindent
The generalized Chaplygin gas (GCG) model allows for an unified description of 
the recent accelerated expansion of the Universe and the evolution of energy 
density 
perturbations. This dark energy - dark matter unification is achieved through 
an exotic background fluid whose equation of state is given by
$p = - A/\rho^{\alpha}$, where $A$ is a positive constant and 
$0 < \alpha \le 1$. Stringent constraints on the model parameters can be
 obtained from recent   WMAP and BOOMERanG 
bounds on the locations of the first few peaks and troughs of the Cosmic
 Microwave Background Radiation (CMBR) power spectrum  as well as SNe Ia data.

\vfill
\newpage

\setcounter{equation}{0}
\setcounter{page}{2}

\baselineskip=20pt

\section{Introduction}

Cosmology is undergoing a blooming period. Precision measurements and
highly predictive theories are coming together to yield a rich lore of
data and methods that scrutinize existing models with increasing
depth. It is quite remarkable that all available data can be fully
harmonized within the {\it Hot Big Bang Model}, an unifying
description in which several branches of physics meet to provide a
consistent and testable scenario for the evolution of the Universe. In
this picture, a particularly relevant role is played by {\it
Inflation}, a period of accelerGiven the potential of the GCG model as
a viable  dark energy-dark matter unification scheme,
 many authors have studied
constraints on  the model parameters from observational data, 
particularly those arising from SNe Ia \cite{Supern}
and gravitational 
lensing statistics \cite{Silva}.

Quite stringent constraints arise also from the study of the position
 of the acoustic peaks and troughs of the CMBR power spectrum. The
 CMBR peaks arise from oscillations of the primeval plasma just before
 the Universe becomes transparent. Driving processes and the ensuing
 shifts on peak positions \cite {Hu}ated expansion in the very early
 Universe that allows for reconciling cosmology with causality and
 leads to a consistent explanation for the origin of the observed
 Large Scale Structure of the Universe. However, in order to fully
 account for the existing observations, one must bring in at least two
 additional new mysteries: the concept of {\it Dark Matter},
 originally proposed to explain the rotation curves of galaxies and
 later used to address the issue of structure formation at large
 scales, and the idea of a smoothly distributed energy that cannot be
 identified with any form of matter, the so-called {\it Dark Energy},
 needed to explain the recently observed accelerated expansion of the
 Universe. Even though these concepts are apparently unrelated, a
 scheme has emerged where an unification of these physical entities is
 possible through the rather exotic equation of state:

\beq
p_{ch} = - {A \over \rho_{ch}^\alpha}~~,
\label{eq:eqstate}
\eeq
\vskip 0.3cm

\noindent
where $A$ a positive constant and $\alpha$ is a constant in the range 
$0 < \alpha \le 1$. This equation of state with $\alpha=1$ was first put 
foward in 1904 by the Russian physicist Chaplygin to describe
adiabatic
 processes
\cite{Chaplygin}; its generalization for $\alpha\neq 1$ was originally
 proposed in Ref.
\cite{Kamenshchik} and the ensuing  cosmology has been analysed in
Ref.
 \cite{Bento1}.
The idea that a cosmological model based on the Chaplygin gas could
lead to
 the  unification
 of dark energy 
and dark matter, thereby reducing two unknown physical entities 
into a single one was first advanced for the case $\alpha=1$ in Refs.
 \cite{Bilic,Fabris}, 
and generalized to $\alpha \neq 1$ in Ref. \cite{Bento1}. 

\section{The Model}

The interesting behaviour of  the equation of state (\ref{eq:eqstate}) can be 
better appreciated by inserting it 
into the relativistic energy-momentum conservation equation, which
implies for
 the 
evolution of the energy density \cite{Bento1}

\beq
\rho_{ch} =  \left(A + {B \over a^{3 (1 + \alpha)}}\right)^{1 \over 1 +
 \alpha}~~,
\label{eq:rhoc}
\eeq 
\vskip 0.3cm

\noindent
where $a$ is the scale-factor of the Universe and $B$ an integration 
constant. Remarkably, this model interpolates between 
a universe dominated by dust and a De Sitter one with an  intermediate
phase  
described by a mixture
of vacuum energy density  and 
 a ``soft'' matter equation of state, $p = \alpha \rho$ ($\alpha \not=
 1$)
 \cite{Bento1}.

Eq.~(\ref{eq:eqstate}) admits, in principle, a wider range of
positive $\alpha$ values; however, the chosen range ensures that 
the sound velocity ($c_s^2 = \alpha A/ \rho_{ch}^{1+\alpha}$) does not exceed,
in the ``soft'' equation of state phase, 
the velocity of light. Furthermore, as pointed out in Ref.~\cite{Bento1}, 
it is only for $0 < \alpha \le 1$ 
that the analysis of the evolution of energy density fluctuations is
physically meaningful.
  
More fundamentally, the model can be described, as discussed in Ref. 
\cite{Bento1},
by a complex scalar field whose action can be written as a generalized 
Born-Infeld
action. This can be seen starting with the Lagrangian density for a
massive 
complex scalar 
field, $\Phi$,
 
\beq
{\cal L} = g^{\mu \nu} \Phi^{*}_{, \mu} \Phi_{, \nu} - V(\vert \Phi \vert^2)~~,
\label{eq:complexfield}
\eeq
\vskip 0.3cm
\noindent
which
can be expressed in terms of its masss, $m$, 
as $\Phi = ({\phi \over \sqrt{2}m} ) \exp(- im \theta)$. Assuming that 
the scale of the inhomogeneities is set by the spacetime variations of
$\phi$
 corresponding 
to scales greater than $m^{-1}$, then
$\phi_{, \mu} << m \phi$,
which, together with Eq.(\ref{eq:eqstate}), leads to a relationship between 
$\phi^2$ and $\rho$:

\beq
\phi^2(\rho_{ch}) = \rho_{ch}^{\alpha} 
(\rho_{ch}^{1 + \alpha} - A)^{{1 - \alpha \over 1 + \alpha}}~~,
\label{eq:phidens}
\eeq
\vskip 0.3cm
\noindent
and  a Lagrangian density that has 
the form of a {\it generalized} Born-Infeld action:

\beq
{\cal L}_{GBI} = - A^{1 \over 1 + \alpha} 
\left[1 - (g^{\mu \nu} \theta_{, \mu} 
\theta_{, \nu})^{1 + \alpha \over 2\alpha}\right]^{\alpha \over 1 + \alpha}~~.
\label{GenBorn-Infeld} 
\eeq
\vskip 0.3cm
Notice that, for $\alpha=1$, one recovers the exact Born-Infeld action.
It is easy to see that Eq.~(\ref{eq:rhoc}) has a bearing  
on the observed
accelerated expansion of the Universe  as it automatically 
leads to an asymptotic phase where the equation of state is dominated by a 
cosmological constant, $8 \pi G A^{1/1+\alpha}$, while at earlier times 
the energy density behaves as if dominated by non-relativistic
matter. This
 dual 
behaviour is at the heart of the unification scheme provided by the GCG model. 
Figure 1 depicts   the way the Universe  evolves in the GCG model.  
It has also been shown that the underlying complex scalar field model
admits,
 under
conditions, 
an inhomogeneous generalization which can be regarded as a unification 
of dark matter and dark energy \cite{Bento1,Bilic} without conflict
 with standard structure formation scenarios \cite{Bento1,Bilic,Fabris,Beca}. 
It is clear that the GCG model collapses into the $\Lambda$CDM model 
when $\alpha = 0$.

These remarkable properties make the 
GCG model an interesting alternative to models where the 
accelerated expansion of the Universe arises from  
an uncancelled cosmological constant  or a rolling scalar field as in
 quintessence
models.

In what follows, we shall discuss the observational bounds that can be
 set on
 the GCG  model
parameters.

\section{Observational Constraints}

Given the potential of the GCG model as a viable  dark energy-dark
matter 
unification scheme,
 many authors have studied
constraints on  the model parameters from observational data, 
particularly those arising from SNe Ia \cite{Supern}
and gravitational 
lensing statistics \cite{Silva}.

Quite stringent constraints arise also from the study of the position
of the acoustic peaks and troughs of the CMBR power spectrum. The CMBR
peaks arise from oscillations of the primeval plasma just before the
Universe becomes transparent. Driving processes and the ensuing shifts
on peak positions \cite {Hu}

\beq 
\ell_{p_m} \equiv \ell_A \left(m - \varphi_m\right)~, 
\label{eq:lm}
\eeq
where $\ell_A$ is the acoustic scale

\beq
\label{eq:la}
l_A = \pi {\tau_0 - \tau_{\rm ls} \over \bar c_s \tau_{\rm ls}}~~,
\eeq $\tau_0$ and $\tau_{\rm ls}$ being the conformal time ($\tau =
\int a^{-1} dt$) today and at last scattering and $\bar{c}_s$ the
average sound speed before decoupling, are fairly independent of post
recombination physics and hence of the form of the potential and the
nature of the late time acceleration mechanism. Hence, the rather
accurate fitting formulae of Ref.~\cite{Doran} can be used to compute
the phase shifts $ \varphi_m$ for the GCG model. In order to calculate
the acoustic scale, we use Eq. (\ref{eq:rhoc}) and write the Universe
expansion rate as
 
\beq
\label{eq:H2}
H^2={8\pi G\over 3}\left[{\rho_{r0}\over a^4}+{\rho_{b0}\over{a^3}}+\rho_{ch0}
\left( A_s + 
{(1-A_s)\over a^{3(1+\alpha)}}\right)^{1/1+\alpha}\right]~~,
\eeq
\vskip 0.3cm
\noindent
where $A_s\equiv A/ \rho_{ch0}^{1+\alpha}$, $\rho_{ch0}\equiv
(A+B)^{1/ 1+\alpha}$ and we have included the contribution of
radiation and baryons as these are not accounted for by the GCG
equation of state. As discussed in Refs.  \cite{Bento4,Bento5}, the
above set of equations allow for obtaining the value of the
fundamental acoustic scale by direct integration, using the fact that
$H^2=a^{-4} \left(d a\over d \tau\right)^2$.

Comparing results from the above procedure with recent bounds on the
  location of the first two peaks and the first trough obtained by the
  WMAP collaboration \cite{WMAP}, namely $\ell_{p_1} = 220.1\pm 0.8,~
  \ell_{p_2} = 546\pm 10,~ \ell_{d_1} = 411.7\pm 3.5$, together with
  the bound on the location of the third peak obtained by the
  BOOMERanG collaboration \cite{Boomerang}, $l_{p3}=825^{+10}_{-13}$,
  leads to quite strong constraints on the model parameters. These
  constraints can be summarized as follows \cite{Bento5}:

\noindent
1) The Chaplygin gas model, $\alpha = 1$, is incompatible
with the data and so are models with $\alpha \gsim 0.6$. 

\vskip 0.3cm
\noindent
2) For $\alpha = 0.6$, consistency 
with data requires for the spectral tilt, $n_s > 0.97$ and $h \lsim 0.68$. 

\vskip 0.3cm
\noindent
3) The $\Lambda$CDM model barely fits the data for  values of the
   spectral tilt
 
$n_s \simeq 1$ 
(notice that  WMAP data leads to  $n_s=0.99\pm0.04$) and for that 
$h > 0.72$ is required. 
For low values of $n_s$, $\Lambda$CDM is preferred to the GCG models
whereas
 for intermediate 
values of $n_s$, the GCG model is favoured only if $\alpha \simeq 0.2$. 

\vskip 0.3cm
\noindent
4) Our study of the peak locations in the $(A_s,\alpha)$ plane shows
that, varying $h$ within the bounds $h=0.71^{+0.04}_{-0.03}$
\cite{WMAP}, does not lead to very relevant changes in the allowed
regions, as compared to the value h=0.71 (see Fig. 3), even though
these regions become slightly larger as they shift upwards for
$h<0.71$; the opposite trend is found for $h>0.71$.

\vskip 0.3cm
\noindent
5) Our results are consistent with the bound found in
Ref. \cite{Bento4} using BOMERanG data for the third peak and Archeops
\cite{Benoit} data for the first peak as well as results from SNe Ia
and age bounds, namely $0.81 \lsim A_s \lsim 0.85$ and $0.2 \lsim
\alpha \lsim 0.6$.

 Bounds from SNe Ia data, which suggest that
$0.6 \lsim A_s \lsim 0.85$ \cite{Supern}, are also consistent with our
 results
 for
$n_s=1$ and $h=0.71$, which yield $0.78\lsim A_s \lsim 0.87$.

\section{Discussion and Outlook}

In this essay, we have described the way the GCG model allows for a
consistent description of the accelerated expansion of the Universe
and purports a scheme for the unification of dark energy and dark
matter. This description is quite detailed and allows for an
unambiguous confrontation with observational data.  For this purpose,
several studies were performed aiming to constrain the parameter space
of the model using Supernovae data, the age of distant quasar sources,
gravitational lensing statistics and the location of the first few
peaks and troughs the CMBR power spectrum, as measured by the WMAP and
BOOMERanG collaborations. These studies reveal that a sizeable portion
of the parameter space of the GCG model is excluded.

More concretely, our results indicate that the Chaplygin gas model,
 $\alpha = 1$, is incompatible with the data and so are models with
 $\alpha \gsim 0.6$.  For $\alpha = 0.6$, consistency with
 observations requires that $n_s > 0.97$.  We find that the
 $\Lambda$CDM model hardly fits the data for $n_s \simeq 1$ and $h >
 0.72$ is required.  For lower values of $n_s$, $\Lambda$CDM is
 preferred to the GCG models whereas for intermediate values of $n_s$
 the GCG model is favoured only if $\alpha \simeq 0.2$.

We conclude that the GCG is a viable dark matter - dark energy model
in that it is compatible with standard structure formation
scenarios. Moreover, although its parameter space is rather
constrained, the model is consistent with all the available
Supernovae, gravitational lensing and CMBR data. Finally, the model
does not suffer from the well-known fine-tuning problems that are
present in alternative dark energy candidate theories such as
$\Lambda$CDM and quintessence models.

\vskip 2cm

\centerline{\bf {Acknowledgments}}

\vskip 0.2cm

\noindent
M.C.B. and  O.B.
acknowledge the partial support of Funda\c c\~ao para a 
Ci\^encia e a Tecnologia (Portugal)
under the grant POCTI/1999/FIS/36285. The work of A.A.S. is fully 
financed by the same grant. 

\vfill



\vfill
\newpage

\begin{figure}[t]
\centering
\leavevmode \epsfysize=12cm \epsfbox{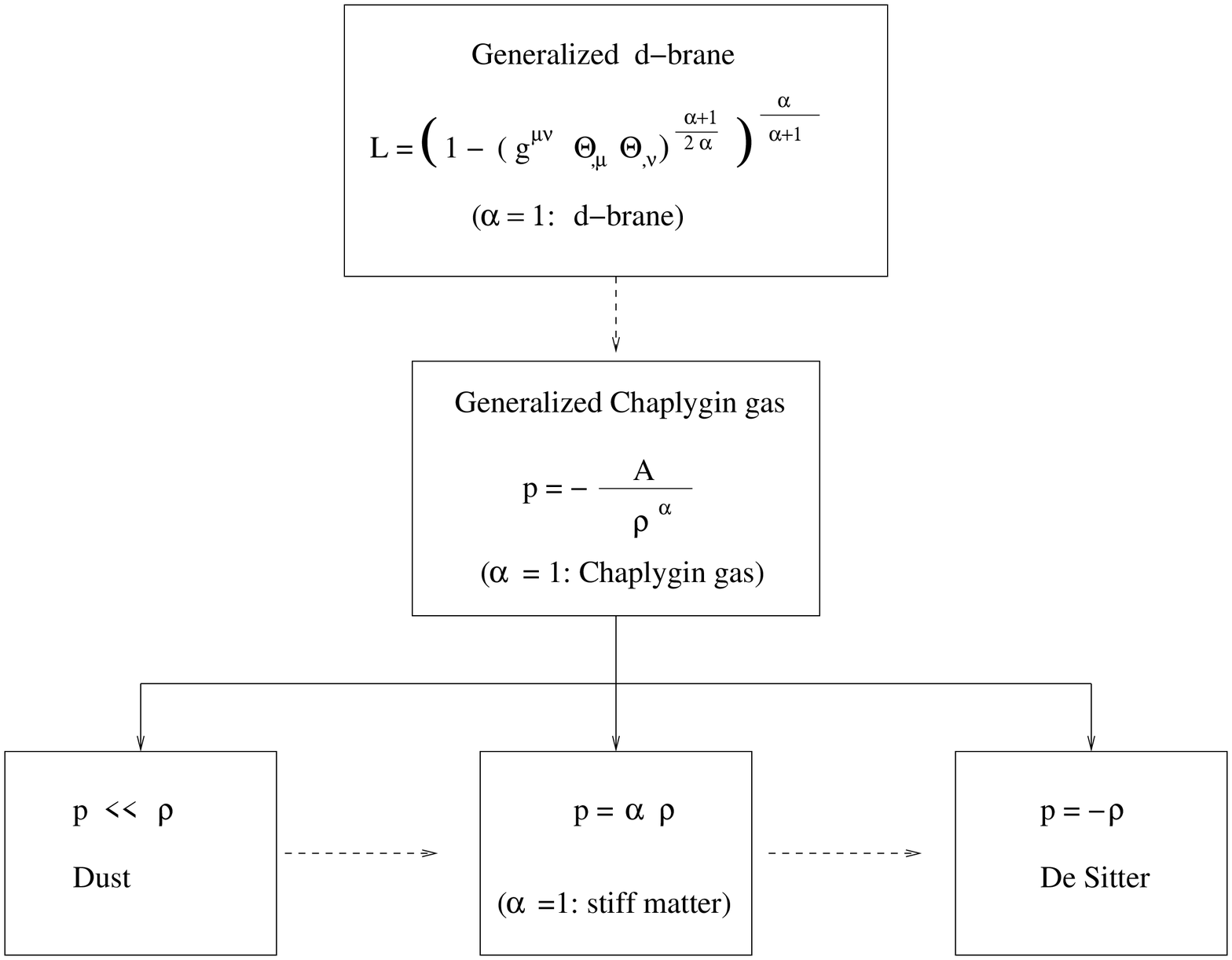}\\
\vskip 0.5cm
\caption{Cosmological evolution of the Universe described by 
the Generalized Chaplygin Gas model.}
\label{comparison2}
\end{figure}

\vfill
\newpage

\begin{figure*}[]
\begin{center}
\includegraphics[height=14cm]{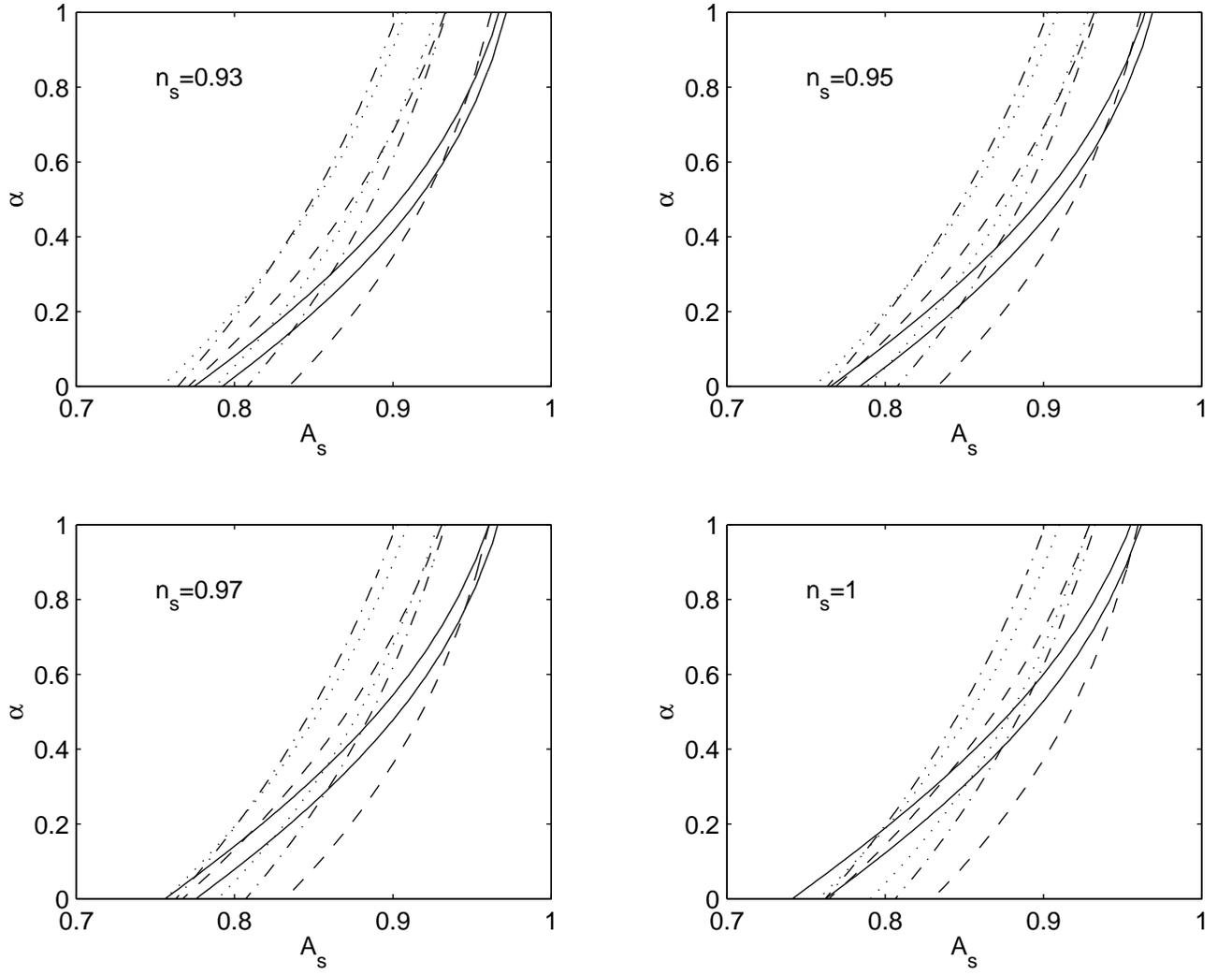}
\caption{\label{fig:h71} Contour plots of the locations of the first three 
 peaks and first trough  of the CMBR power spectrum, in the $(A_s,
 \alpha)$
 plane, for 
a GCG model, with $h=0.71$, for different values of $n_s$.
Full, dashed, dot-dashed and dotted contours correspond to observational
bounds on 
$\ell_{p_1}$, $\ell_{p_2}$, $\ell_{p_3}$ and $\ell_{d_1}$, respectively.}
\end{center}
\end{figure*}

\end{document}